\documentclass{aa}  
\usepackage{natbib}
\usepackage{xspace}
\usepackage{caption}
\usepackage{soul}

\bibpunct{(}{)}{;}{a}{}{,}

\usepackage[dvipsnames]{xcolor}

\newcommand{\Obelisk}{\textsc{Obelisk}\xspace}
\newcommand{\hi}{\ion{H}{I}\xspace}
\newcommand{\hii}{\ion{H}{II}\xspace}
\newcommand{\hei}{\ion{He}{I}\xspace}
\newcommand{\heii}{\ion{He}{II}\xspace}

\newcommand{\lya}{\ensuremath{\text{Lyman-}\alpha}\xspace}

\newcommand{\fescLyC}{\ensuremath{f_{\text{esc}}(\text{LyC})}\xspace}
\newcommand{\fescLya}{\ensuremath{f_{\text{esc}}(\text{Ly}\alpha)}\xspace}
\newcommand{\Msun}{\ensuremath{M_{\odot}}\xspace}
\newcommand{\Muv}{\ensuremath{M_{\rm UV}}\xspace}

\newcommand{\code}[1]{\textsc{\MakeLowercase{#1}}}

\usepackage{graphicx}
\usepackage{txfonts}
\usepackage[colorlinks=true,allcolors=blue]{hyperref}
\begin{document}

   \title{Modeling LAEs in the epoch of reionization with OBELISK
}
\titlerunning{Modeling LAEs in the epoch of reionization with OBELISK}

   \subtitle{The connection between Lyman-$\alpha$ spectra and Lyman-continuum
escape}

   \author{Emma Giovinazzo
          \inst{1}
          \and
          Maxime Trebitsch \inst{2} 
          \and Valentin Mauerhofer \inst{2}
          \and Pratika Dayal \inst{2}
          \and Pascal A. Oesch \inst{1}$^,$\inst{3}
          }

   \institute{Department of Astronomy, University of Geneva, Chemin Pegasi 51, 1290 Versoix, Switzerland
         \and
             Kapteyn Astronomical Institute, University of Groningen, P.O. Box 800, 9700 AV Groningen, The Netherlands
             \and 
             Cosmic Dawn Center (DAWN), Niels Bohr Institute, University of Copenhagen, Jagtvej 128, K\o benhavn N, DK-2200, Denmark
             }

   \date{Received ---; accepted ---}

 
  \abstract
   {\lya{} emitters (LAEs) are particularly useful objects in the study of the epoch of reionization. \lya{} profiles can be used to estimate the number of ionizing photons that are able to escape galaxies, and therefore to understand which objects contributed to reionization. However, \lya{} is a resonant line and its complex radiative transfer effects make the interpretation of the line challenging and require the use of appropriate radiative transfer methods for anything but the simplest gas distributions, such as uniform gaseous spheres, slabs, or cubes.}
   {With this work, we aim to study the properties of simulated LAEs, and the robustness of these inferred properties during a change in the dust model. We also explore the Lyman continuum (LyC) escape fraction of these galaxies and compare our results with observationally calibrated methods to infer this quantity from the \lya{} spectrum.}
   {We used the radiative transfer code \code{rascas} to perform synthetic observations of 13 flux-selected galaxies from the \Obelisk simulation at a redshift of $z=6$, toward the end of the epoch of reionization. Each galaxy was observed in \lya{}, as well as ionizing and nonionizing continuum from 48 different viewing angles.}
   {We show that the \lya{} profiles emitted from a galaxy present large variations with a change in viewing angle and that the relation between peak separation and the \lya{} escape fraction is not as strong as previously found, as we find lines of sight with both a low peak separation and a low escape fraction, due to their dust content. We also show that the properties of the \lya{} line are reasonably robust during a change in the dust model. Lastly, we compare the LyC escape fractions that we derive from the simulation to three observationally calibrated methods of inferring this quantity. We determine that none of these relations reproduce the scatter that we find in our sample, and that high escape fraction lines of sight have both a low peak separation and a low dust extinction in the ultraviolet (UV).}
   {}

   \keywords{galaxies: high-redshift --
            reionization  --
                 Radiative transfer --
                Line: profiles
               }

   \maketitle
   
%

\section{Introduction}

The epoch of reionization (EoR) marks the last phase transition of the Universe, during which the initially cold and neutral intergalactic medium (IGM) became ionized. 
This period also marks the formation of the first luminous objects, which formed at the center of collapsed halos and produced sufficient amounts of ionizing radiation to drive the reionization process (e.g., \citealp{Barkana_2001}, \citealp{Dayal_2018}). 

Current constraints from quasar spectra place the end of the EoR around $z\sim5.5-6$ \cite[e.g.,][]{Wyithe_2005, Fan2006, Eilers_2018, Schroeder_2012,Becker_2021,Bosman22}, but it is still unclear which sources are responsible for the emission of the majority of the ionizing photons.
It is commonly believed that the main driver of the reionization of hydrogen in the Universe is massive star formation in galaxies \citep[e.g.,][]{Robertson_2015, Bouwens15, Finkelstein19, Naidu20, Yung_2020b, Yung_2020a, trebitsch2022reionization}, as young and massive stars emit copious amounts of Lyman continuum (LyC) radiation ($\lambda<912$ \AA). This scenario implicates a very patchy process \citep[e.g.,][]{D_Aloisio_2015, Davies_2016, Eilers_2018}, whereby galaxies are able to ionize gas in their surroundings, forming growing bubbles of ionized gas around themselves until these bubbles overlap and the entirety of the IGM is ionized. 
Another potentially important source of ionizing photons are active galactic nuclei (AGNs)  \citep[e.g.,][]{Madau15,Giallongo15}. 
However, studies of the AGN luminosity function in the EoR have shown that AGNs are only minor contributors of reionization and fail to solely reionize the universe by $z \sim 6$ \cite[e.g.,][]{Onoue_2017,Parsa_2017, Yung_2021, Matsuoka_2023}, although according to theoretical models, AGNs could dominate the ionizing photon production at the highest masses \citep{Dayal_2020}.

To properly understand this process, it is necessary to determine which galaxies leak the most ionizing photons, which depends on large-scale properties of the IGM such as clumping, and on properties of the galaxies such as their star formation rates (SFRs) and the escape fraction of LyC photons \fescLyC. This last parameter quantifies the fraction of ionizing photons that are able to escape the galaxy, and therefore ionize the surrounding medium. Determining this quantity through observations is challenging, as it is only possible to observe the escaped radiation. Only recently have surveys been able to accumulate large numbers of directly detected escaping ionizing photons (e.g., \citealp{Bergvall_2006}, \citealp{Shapley_2006}, \citealp{Heckman_2011}, \citealp{Leitet_2013}, \citealp{Izotov_2016a, Izotov_2016b}, \citealp{Izotov_2018a, Izotov_2018}, \citealp{Fletcher_2019}, \citealp{Davis_2021}, \citealp{Izotov_2021}, \citealp{Flury_2022a, Flury_2022b}, \citealp{Saxena_2022b}). Moreover, it is impossible to detect LyC photons at high redshifts, $z \gtrsim 4$, as they are completely absorbed by the residual \hi in the IGM. It is therefore necessary to use indirect indicators that link \fescLyC to observable properties, which are often calibrated on low-redshift observations.
Some previously studied indicators are based on metal lines and line ratios, such as the \ion{Mg}{II} line \citep[e.g.,][]{Chisholm_2020, Xu_2022}, the \ion{C}{IV} line \citep[e.g.,][]{Saxena_2022, Schaerer_2022}, the $\ion{O}{III}\lambda5007$/$\ion{O}{II}\lambda3727$ line ratio ($O_{32}$) \citep[e.g.,][]{Jaskot_2013,Nakajima_2014, Izotov_2018, Paalvast_2018, Tang_2021}, or also the ultraviolet (UV) $\beta$ slope \citep[e.g.,][]{Chisholm_2022, Flury_2022b}.

Another notable indicator is the \lya{} line \citep[e.g.,][]{Verhamme_2015, Verhamme_2016, Izotov_2021}. Both \lya{} and LyC are expected to escape galaxies from the same low neutral hydrogen column density paths \citep[e.g.,][]{Dijkstra_2014, Verhamme_2015, gazagnes_2020, Kakiichi_2021, begley2023connecting}, making this line a very interesting indicator. The most promising characteristic of this line for the estimation of \fescLyC is the peak separation ($v_{\rm sep}$), first suggested by \cite{Henry_2015} and \cite{Yang_2017} and then confirmed by \cite{Izotov_2018} and the Low-Redshift Lyman Continuum Survey (LzLCS) \citep{Flury_2022b}.

Galaxies that are particularly bright in the \lya{} line, due to their star-forming nature \citep{Dijkstra_2014}, are called \lya{} emitters (LAEs). The LAEs have been observed and studied at low redshift \citep{Orlitova_2018}, although they are quite rare in the nearby universe \citep{Hayes_2015}. They are particularly interesting when used to study the EoR, both as probes of ionized bubbles in the partially neutral IGM and as contributors of reionization \citep{Hu_2016, Matthee_2018, Songaila_2018, Meyer_2020}. The resonant nature of \lya{} does, however, add a layer of complexity to the study of this line in the EoR, as \lya{} photons are unlikely to escape the neutral intergalactic gas. A relevant review of LAEs has been produced by \citet{Ouchi_2020}.

In this work, we analyze synthetic observations of a sample of LAEs extracted from the \Obelisk cosmological simulation \citep{Trebitsch_2021} in order to study their \lya properties in connection with the escape fraction of ionizing photons. In particular, we study how the modeling of the dust distribution inside the galaxies affects the \lya transfer, and how this, in turn, impacts the usability of \lya diagnostics in the study of the sources of reionization. In this work, we define \fescLyC as the absolute escape fraction along the observed line of sight.

This work is structured as follows. In Sect.~\ref{ch:methods}, we present the \Obelisk simulation and the radiative transfer code used in the \lya{} and LyC post-processing. In Sect.~\ref{ch:Lya_properties}, we show the properties of the synthetic \lya{} profiles. Then, in Sect~\ref{ch:dust_model_comp}, we study the impact of the dust modeling on our results. In Sect.~\ref{ch:Estimate_lyc}, we compare the LyC escape fraction, \fescLyC, directly measured in the simulation to standard estimates used throughout the literature. 

\section{Methods} \label{ch:methods}

\subsection{\Obelisk simulation}

In this paper, we analyze a sample of galaxies extracted from the \Obelisk cosmological simulation \citep{Trebitsch_2021}, which re-simulates at a much higher resolution and down to $z \simeq 3.5$ the evolution of a large, overdense region ($V \simeq 23\,h^{-3}\mbox{cMpc}^3$) within the \textsc{Horizon-AGN} \citep{Dubois_2014} $(100\,h^{-1}\,\mbox{Mpc})^3$ cosmological volume. In this section, we highlight the features of the simulation most relevant to the study of LAEs and refer the reader to \citet{Trebitsch_2021} for further details.

The simulation uses the same WMAP-7 \citep{Komatsu_2011} $\Lambda$CDM cosmology as \textsc{Horizon-AGN}, with the cosmological parameters: the Hubble constant, $H_0 = 70.4\,\mbox{km}\,\mbox{s}^{-1}\mbox{Mpc}^{-1}$, the dark energy density parameter, $\Omega_\Lambda = 0.728$, the matter density parameter, $\Omega_m = 0.272$, the baryon density parameter, $\Omega_b = 0.0455$, amplitude normalization of the power spectrum, $\sigma_8 = 0.81$, and the spectral index, $n_s = 0.967$.
The high-resolution region was defined in the initial conditions as the convex hull enclosing all the particles that will end up within a sphere of radius $\approx 2.5\,h^{-1}\,\mbox{cMpc}$ centered on the most massive halo in the \textsc{Horizon-AGN} simulation at $z=2$. In this region, dark matter (DM) particles have a mass resolution of $1.2\times 10^6\,\Msun$, and gas cells are allowed to be refined up to $35\,\mbox{pc}$ if the total mass in the cell exceeds eight times the mass resolution.

\Obelisk was run with the \textsc{Ramses-RT} \citep{Rosdahl_2013,Rosdahl_2015} radiation-hydrodynamics module of the adaptive mesh refinement (AMR) code \textsc{Ramses} \citep{Teyssier_2002}, which follows the evolution of DM, gas, stars, black holes, and the (ionizing) radiation field.
Gas in the simulation is described as a monoatomic gas with an adiabatic index of $\gamma = 5/3$ and was evolved using an unsplit second-order MUSCL-Hancock scheme \citep{vanLeer1979} with the HLLC Rieman solver \citep{Toro_1994}. The masses of DM and star particles were projected onto the grid using a cloud-in-cell interpolation (with the DM particles projected on a coarser grid with $\Delta x \simeq 540\,\mbox{pc}$) and combined with the gas density to evolve the gravitational field.
The radiation field is split into three frequency intervals corresponding to \hi, \hei, and \heii-ionizing photons. The ionizing radiative transfer used in \Obelisk then follows the standard \textsc{Ramses-RT} methodology, using a first-order Godunov method with the M1 closure \citep{1984JQSRT..31..149L,1999CRASM.329..915D} to evolve the radiation field. As is described in \citet{Trebitsch_2021}, we used the variable speed of light approximation of \citet{Katz_2017}, which uses the reduced speed of light approximation in regions resolved better than $\Delta x \simeq 2\,\mbox{kpc}$ (this is the case for all the intra-halo gas in this study). While \Obelisk separately tracks the contribution of massive stars and AGNs to the radiation field, we treat the two components as one in this study.
Finally, the gas thermodynamical evolution was followed using the nonequilibrium hydrogen and helium thermochemistry method from \textsc{Ramses-RT}, described in \citet{Rosdahl_2013}, which includes a cooling contribution of metals down to a floor temperature of $50\,\mbox{K}$. Importantly for this work, we followed on the fly the ionization state of hydrogen and helium, which is necessary to accurately predict the \lya emissivity from the gas.

In the simulation, star particles are formed with a mass of $m_\star \simeq 10^4\,\Msun$ in the dense and turbulent ISM (above a number density of $n_{\rm SF} = 5\,\mbox{cm}^{-3}$ and a turbulent Mach number of $M\geq 2$), assuming a local star formation efficiency per free-fall time, $\epsilon_\star$, defined using the thermo-turbulent model presented, for example, in \citet{Trebitsch_2017} and \citet{Kimm_2017}.
After they are formed, star particles emit ionizing radiation following the \textsc{Bpass} v2.2.1 binary stellar population model \citep{Eldridge_2017, Stanway_2018}, as a function of the age and metallicity of the particle.
After a delay of $t_{\rm SN} = 5\,\mbox{Myr}$, a mass fraction, $\eta_{\rm SN} = 0.2$, of the initial stellar population in each particle explodes as supernovae, injecting mass, energy, momentum, and metals into their environments. Supernova feedback was modeled following the mechanical feedback implementation of \citet{Kimm_2014, Kimm_2015}, with an additional boost due to the preprocessing of the ISM in (unresolved) \hii regions inspired by the results of \citet{Geen_2015}.

The simulation also includes a detailed model for black hole formation, growth through accretion and mergers, a dual-mode AGN feedback model, and follows ionizing radiation resulting from accretion. In this work, we focus on galaxies less massive than $\lesssim 10^{10}\,\Msun$, where the black hole growth is inefficient (see for example \citealt{Habouzit_2017}). As a result, we ignore the AGN effect on the galaxies: for more details on the black hole model, we refer the reader to \citet{Trebitsch_2021}.

Importantly for this work, \Obelisk models the evolution of dust separately from the metals, such that the dust-to-metal ratio is allowed to vary locally. Dust is released in the ISM in supernovae, and the dust content in a cell grows via the accretion of metals from the gas phase. Conversely, dust is destroyed in supernova shocks and through thermal sputtering in hot gas. While the dust is not coupled with the radiative transfer or with the cooling in the \Obelisk simulation, it plays an important role in setting the \lya properties of galaxies. We describe how we use this spatially varying dust content to estimate the attenuation of \lya and continuum photons in Section~\ref{sec:obelisk_dust}.

Finally, galaxies (and their DM haloes) were identified using the \textsc{AdaptaHOP} algorithm \citep{Aubert_2004}, operating on all collisionless particles at the same time (DM and stars). Only galaxies with more than 100 star particles and 100 DM particles are considered in this work: as such, the lowest possible stellar mass is $\simeq 10^{6}\,\Msun$. For all galaxies, the (dust-attenuated) UV magnitude was computed by casting 192 rays isotropically from each star particle, integrating the dust column density along these lines of sight, and computing the direction-averaged total UV luminosity. We used these UV magnitudes to define our galaxy sample, as is described in the next section.

\subsection{Galaxy sample}

We selected galaxies in order to create a synthetic flux-selected sample targeting faint Lyman-break galaxies corresponding to typical LAEs at the end of the EoR. For this, we focused on the $z = 6$ snapshot of the \Obelisk simulation and selected all (central) galaxies with a 3D-averaged, dust-attenuated UV magnitude in the range of $-19.5 \leq \Muv \leq -19$. Because different galaxies have different levels of attenuation, this corresponds to a much broader range of intrinsic UV luminosities, with magnitudes between $-19.4$ and $-23.4$.
We summarize the properties of our sample in Table~\ref{Tab:table_galaxies}.
\begin{center}
\begin{table*}[]
{\centering
\begin{tabular}{l|l|l|l|l|l|l|l}
ID & $M_{*}$ [$M_{\odot}$] & SFR [$M_{\odot}$/yr] & $M_{\mathrm{UV}}^{\mathrm{obs}}$ & $M_{\mathrm{UV}}^{\mathrm{intr}}$ & $f_{\mathrm{esc}}(\mathrm{LyC})$ & $M_{\textrm{dust}}$ [$M_{\odot}$] \\ \hline
 2899 & 1.4 $\cdot 10^{10}$ & 31.0 & -19.4 & -21.5 &0.001 & 1.1 $\cdot 10^{7}$\\
 24688 & 6.4 $\cdot 10^{9}$ & 35.7 & -19.4 & -23.4 &0.004 & 8.7 $\cdot 10^{6}$\\
 55468 & 4.9 $\cdot 10^{9}$ & 17.9 & -19.0 & -21.9 &0.002 & 9.8 $\cdot 10^{6}$\\
 36112 & 3.7 $\cdot 10^{9}$ & 22.8 & -19.4 & -22.3 &0.007 & 1.4 $\cdot 10^{7}$\\
 41163 & 3.2 $\cdot 10^{9}$ & 8.4 & -19.0 & -21.9 &0.007 & 5.5 $\cdot 10^{6}$\\
 3639 & 2.6 $\cdot 10^{9}$ & 18.1 & -19.1 & -21.8 &0.001 & 5.9 $\cdot 10^{6}$\\
 42655 & 1.7 $\cdot 10^{9}$ & 10.0 & -19.2 & -21.5 &0.023 & 5.8 $\cdot 10^{6}$\\
 8776 & 1.6 $\cdot 10^{9}$ & 6.2 & -19.0 & -20.9 &0.026 & 8.3 $\cdot 10^{6}$\\
 39021 & 1.5 $\cdot 10^{9}$ & 9.5 & -19.0 & -21.1 &0.033 & 8.0 $\cdot 10^{6}$\\
 76168 & 1.5 $\cdot 10^{9}$ & 8.0 & -19.2 & -21.6 & 0.011 & 3.1 $\cdot 10^{6}$\\
 10345 & 2.5 $\cdot 10^{8}$ & 1.5 & -19.1 & -20.0 &0.235 & 1.5 $\cdot 10^{6}$\\
 57172 & 2.4 $\cdot 10^{8}$ & 1.5 & -19.3 & -19.8 &0.485 & 1.2 $\cdot 10^{6}$\\
 67244 & 7.0 $\cdot 10^{7}$ & 0.7 & -19.0 & -19.4 &0.195 & 2.1 $\cdot 10^{5}$
\end{tabular}\par
}
\caption{Characteristics of the 13 galaxies in the sample. Here, $M_{*}$ is their stellar mass, SFR is the star formation rate, calculated over 10 Myr, $M_{\mathrm{UV}}^{\mathrm{obs}}$ is the observed absolute dust-attenuated UV magnitude, $M_{\mathrm{UV}}^{\mathrm{intr}}$ is the intrinsic absolute UV magnitude, \fescLyC is the LyC escape fraction, and $M_{\textrm{dust}}$ is the dust mass.}\label{Tab:table_galaxies}
\end{table*}
\end{center}
\subsection{Ly$\alpha$ emission and transfer}

The radiative transfer of Lyman-$\alpha$ photons through our galaxy sample was performed in post-processing with the 3D Monte Carlo radiative transfer code \code{rascas} \citep{Michel_Dansac_2020}. We used the standard ``peeling-off'' technique \citep[e.g.,][]{Yusef-Zadeh_1984,Zheng_2002, Whitney_2011} to produce synthetic observations of the \lya line on the fly. At each scattering, both the probability that the photon would have gone in the direction of the observer and the probability that the photon would have escaped are calculated. This method allows for the production of synthetic spectra and images in any direction and is performed at the same time as the radiative scattering. 

The first step in the \lya transfer is the emission of Monte Carlo photons, each representing a fraction of the total \lya flux. This was done by computing the \lya{} emissivity directly from the cells in the simulation, accounting both for recombination and collisional excitation mechanisms. The emitting gas was then sampled using $10^5$ photons, following the luminosity distribution of the cells.
In a given cell, the number of photons emitted due to recombination per unit time was computed as
\begin{equation}
    \dot N_{\gamma, rec} = n_e n_p \epsilon^B_{Ly\alpha}(T)\alpha_B (T) \times (\Delta x)^3
,\end{equation}
where $n_e$ and $n_p$ are the electron and proton number densities, and $(\Delta x)^3$ is the cell volume. The $B$ stands for case B recombination, with $\alpha_B$ being the recombination coefficient given by \citep{Hui_1997}:
\begin{equation}
    \alpha_B = 2.753 \times 10^{-14} \textrm{cm}^3 \textrm{s}^{-1} \frac{\lambda^{1.5}_{\hi}}{(1+(\lambda_{\hi}/2.740)^{0.407})^{2.242}}
,\end{equation}
and $\epsilon^B_{Ly\alpha}(T)$ being the fraction of recombination producing \lya{} photons. We evaluated it using the fit from \citet{Cantalupo_2008}, which is based on the tabulated values from \citet{Pengelly_1964} for $T>10^3$ K and \citet{Martin_1988} for $T<10^3$ K:
\begin{equation}
    \epsilon^B_{Ly\alpha}(T) = 0.686 - 0.106 \log(T_4) - 0.009 \times (T_4)^{-0.44}
.\end{equation}
Instead, the number of \lya{} photons emitted from the gas per unit time in each cell due to collisions is given by
\begin{equation}
    \dot N_{\gamma, col} = n_e n_{\hi} C_{Ly\alpha}(T) \times (\Delta x)^3
,\end{equation}
where $n_{\hi}$ is the number density of neutral hydrogen atoms and $C_{Ly\alpha}$ is the rate of collisional excitation from level $1s$ to $2p$, given by \citep{Goerdt_2010}:
\begin{equation}
    C_{Ly\alpha}(T) = \frac{2.41 \times 10^{-6}}{T^{0.5}} \left(\frac{T}{10^4}\right)^{0.22} \times \textrm{exp} \left( -\frac{h \nu_{Ly\alpha}}{kT} \right) \textrm{cm}^3 \textrm{s}^{-1}.
\end{equation}
The Monte Carlo photons are then placed in a random position within their emission cell and emitted in a direction randomly drawn from an isotropic distribution. The initial wavelength is taken from a Gaussian distribution centered around the \lya frequency in the rest-frame of emitting gas, with a width following the thermal velocity dispersion of the gas.

Once the photons have been emitted, they travel until their next scattering event, which can be either on \hi or on a dust grain. In practice, the optical depth to the next scattering event is drawn from $\tau_{\mathrm{event}} = -\mathrm{ln}(r)$, where $r$ is a random number between zero and one. The photons are then propagated until they reach this optical depth. In each cell, the total optical depth (gas + dust) is given by
\begin{equation}
    \tau_{\mathrm{tot}} = r(n_{\hi}\sigma_{\hi} + n_d \sigma_d) = \tau_\hi + \tau_d
.\end{equation}
The neutral hydrogen cross section is defined as
\begin{equation}
    \sigma_{\hi} = f_{12} \frac{\pi e^2}{m_e c \Delta \nu_D} \phi(x)
,\end{equation}
where $f_{12}$ is the \lya oscillator strength, and $\phi(x)$ is the (Voigt) line profile. We discuss in Sect.~\ref{sec:dustmodels} the details of the evaluation of the dust optical depth, $\tau_d$.
We propagated photons until they were either destroyed by a dust grain or escaping the galaxy, defined here as crossing the virial radius, $R_{\rm vir}$, of the halo.


\subsection{Mock observation}\label{ch:mock_obs}

Each galaxy was observed along 48 different lines of sight, uniformly distributed around a sphere, selected using the Hierarchical Equal Area isoLatitude Pixelization (HEALPIx) decomposition (\citealt{Gorski_2005}, \citealt{Calabretta_2007}), which divides a sphere into equal area pixels. 
While none of these directions are strictly independent, previous works (e.g., \citealp{blaizot2023simulating}, but see also \citealp[]{Verhamme_2012,Behrens2014,Smith_2019}) have shown that the \lya line emerging from galaxies is highly anisotropic and can mimic a large range of observed spectra. We used this characteristic to our advantage by treating each line of sight as a separate point.

For each sight line, \code{rascas} outputs a data cube with two spatial directions and one spectral direction, from which we can extract the spectra by collapsing the cube in the spatial directions. The extracted \lya{} spectra are observed in a rest-frame wavelength range of 1210 \AA{} to 1220 \AA{} in 250 pixels. This corresponds to a wavelength resolution of 0.04 \AA{} or $\sim$10 km $\mathrm{s}^{-1}$. These profiles can then be shifted to the observer frame and analyzed. 
To identify the peaks of the profiles, we used the \texttt{findpeaks} python module \citep{Taskesen_findpeaks_is_for_2020}.
We first reduced the Monte Carlo noise by applying a Gaussian smoothing with a sigma of two pixels to the profile. Smoothing on this scale should ensure that we do not remove any significant peaks. We then applied the \texttt{findpeaks} module on this new profile, to identify all the local maxima in the distribution. Next, we computed the noise level as the standard deviation of our unsmoothed data when subtracted to a polynomial fit to the profiles, which were smoothed over four pixels to perform the fit. We used the noise to discern between significant peaks and peaks resulting from noise, by only selecting from all the peaks only those with a signal-to-noise ratio above three. The peak separation was then computed as the separation between the identified peaks. If only one peak could be identified, the peak separation was set as 0, and in the case of triple peaks we determined the peak separation to not be identifiable without a visual inspection so we removed the spectrum from the peak separation dataset. Overall, we find 277 spectra where only one peak could be identified, 44\% of the total sight lines, and only nine spectra where more than two peaks were identified, which is only 1.4\% of the total dataset.
For all other uses, the spectra still need to be smoothed to reduce the Monte Carlo noise, but that was done directly with the \code{rascas} function, which also loads the mock observations and shifts them into the observer frame. This function takes as the Gaussian sigma a wavelength value, which we chose as 0.5 \AA{}. This level of smoothing is also necessary to produce spectra that can be comparable to observations, like those seen in Figure~\ref{fig:all_spectra}. 

Finally, we performed three additional \code{rascas} runs to measure the ionizing and nonionizing UV luminosities and estimate the continuum around \lya{} for each galaxy along the same lines of sight. For the nonionizing UV, we propagated photons in the wavelength range 1480 \AA{} to 1520 \AA{} and only considered dust absorption using the same approach as for \lya (see Section~\ref{sec:dustmodels} for details).
The ionizing photon run considers both gas and dust absorption and was only used to investigate the variation in \fescLyC with direction.
To estimate the continuum at the \lya{} wavelength, we sampled the emission from stars with $10^5$ photons, using the tabulated spectral shape, between the wavelengths of 1150 \AA{} to 1200 \AA{}.
In all three cases, we obtained data cubes from which we extracted the spectra by collapsing the cube and smoothing all the spectra with the same Gaussian, and we then calculated the luminosity and flux, just as was done with \lya{}.

\subsection{Dust models}
\label{sec:dustmodels}

We now focus on the calculations required to assess the dust optical depth, $\tau_d$, in the \lya transfer.
Since \Obelisk includes a model for dust evolution, we can take two different approaches to estimate $\tau_d$ in this work. First, we directly make use of the \Obelisk dust model. Then, we also compute $\tau_d$ following the formulation of \cite{Laursen_2009}, which is a relatively standard approach and is the default model used in \code{rascas}.

\subsubsection{Dust from \Obelisk}
\label{sec:obelisk_dust}

The first model of dust attenuation that we considered
relies on the dust distribution tracked by the \Obelisk simulation: the dust mass density, $\rho_d$, and the dust-to-metal ratio, $\mathcal{D}/Z = \rho_d/\rho_Z$, is therefore allowed to vary within a galaxy.

However, the knowledge of $\rho_d$ is insufficient to estimate the dust optical depth: indeed, the dust cross section depends on the grain size, which is not tracked directly in \Obelisk.\footnote{\label{fn:dustsize}While the dust evolution model implicitly assumes that all grain have an average size of $a = 0.1\mu\mbox{m}$, this is not appropriate to derive an extinction curve.} Instead, we computed the dust optical depth over a path of length $L$ directly as $\tau_d = \alpha^{\rm ext} L = \rho_d \kappa_d L$, where $\kappa_d$ is the absorption coefficient and 
\begin{equation}
    \label{eq:alpha_lambda}
    \alpha^{\rm ext}(\lambda) = \int_{a_{\min}}^{a_{\max}} \pi a^2 \sum_{i=s,c} n(a) Q_{\rm ext}^i (a, \lambda)da.
\end{equation}
Here, $Q_{\rm ext}$ is the extinction efficiency factor as a function of grain size and wavelength taken from \cite{Weingartner_2001} and \cite{Laor_1993} for two grain types (silicates, $s$, and carbonaceous grains, $c$) and $n(a)$ is the number density of grains of size $a$. We assume that silicates have a mass fraction of 54\% and 46\% for the carbonaceous grains \citep{2009MNRAS.394.1061H}.
We then converted the mass density into a number density by assuming that the grain size is well described by the MRN distribution \citep{1977ApJ...217..425M}, $dn/da = C a^{-3.5}$, where $C$ is a proportionality constant. We note that $n_d$ is the total number density, such that $n(a) = n_d C a^{-3.5}$. For a dust grain specific density, $\mu_g$, we can write the mass density as \begin{equation*}
    \rho_d = \int_{a_{\min}}^{a_{\max}} \frac{4}{3}\pi a^3 \mu_g C n_d a^{-3.5} da = \frac{8}{3}\pi \mu_g n_d C \left(\sqrt{a_{\max}} - \sqrt{a_{\min}}\right)
\end{equation*}
for grain sizes between $a_{\rm min}$ and $a_{\rm max}$, and so
\begin{equation*}
    n(a) = \frac{3\rho_d}{8\pi \mu_g  \left(\sqrt{a_{\max}} - \sqrt{a_{\min}}\right)} a^{-3.5}
.\end{equation*}

When evaluating the dust optical depth, \code{rascas} directly uses the form $\tau_d = \rho_d \kappa_d L$. We therefore rewrote Eq.~\ref{eq:alpha_lambda} as
\begin{equation}
    \label{eq:kappa_d}
    \kappa_d = \frac{3}{8\mu_g  \left(\sqrt{a_{\max}} - \sqrt{a_{\min}}\right)} \int_{a_{\min}}^{a_{\max}} a^{-1.5} \sum_{i=s,c} Q_{ext}^i (a, \lambda)da.
\end{equation}
We integrated Equation ~\ref{eq:kappa_d} numerically to obtain $\kappa_d(\lambda)$, and obtained a fit (used for numerical efficiency) using the python package LMFIT \citep{newville_2015}, with a relative difference to the tabulated values lower than 5\% around the \lya wavelength.

\subsubsection{Dust following the metals}

The second model that we considered is the default dust model used in \code{rascas} that assumes a dust-to-metal ratio proportional to the ionization fraction of the cell. It has been used previously in, for example, \citet{blaizot2023simulating} to post-process simulations where the dust is not tracked self-consistently. Specifically, we used the Large Magellanic Cloud (LMC) model of \citet{Laursen_2009}, based on previous works by \cite{Pei_1992} and \cite{Gnedin_2008}.

The dust optical depth along a path of length $L$ in a cell was computed as $\tau_d = n_d \sigma_d L$, where $n_d$ is a dust ``pseudo-density'' and $\sigma_d$ is the dust cross section per hydrogen atom: the dust (pseudo-)density, $n_d$, therefore gives the effective number of hydrogen atoms required to reach a given optical depth.
The dust density is given by
\begin{equation}
    n_d = (n_{\hi} + f_{ion}n_{\hii}) \frac{Z}{Z_{0}}
,\end{equation}
where $f_{\mathrm{ion}} = 0.01$ is the fraction of dust grain surviving in the ionized gas, $Z$ is the local metallicity of the gas, and $Z_0$ is a reference metallicity, here taken to be that of the LMC. Our galaxies are therefore assumed to have an LMC-like dust-to-metal ratio, roughly proportional to the neutral hydrogen fraction of each cell.
The cross section, $\sigma_d$, per hydrogen atom is given, close to the \lya frequency, by \citep{Laursen_2009}:
\begin{equation}
    \sigma_d/10^{-21} \textrm{cm}^2 = 0.723 + 4.46 \times 10^{-5} (T/10^4 \textrm{K})^{1/2} x.
\end{equation}
Here, $x$ is the parametrization of the frequency, $\nu$, defined as $x \equiv (\nu - \nu_0)/\Delta \nu_{\mathrm{D}}$, where $\nu_0$ is the line center and $\nu_{\mathrm{D}}$ is the Doppler width of the line.

\subsection{Dust model comparison}\label{ch:dust_model_comp}

In this section, we compare the two different dust models considered in this work to assess how they affect the predicted observable properties of our simulated galaxies.

\begin{figure}
    \centering
    \includegraphics[width = 0.95 \linewidth]{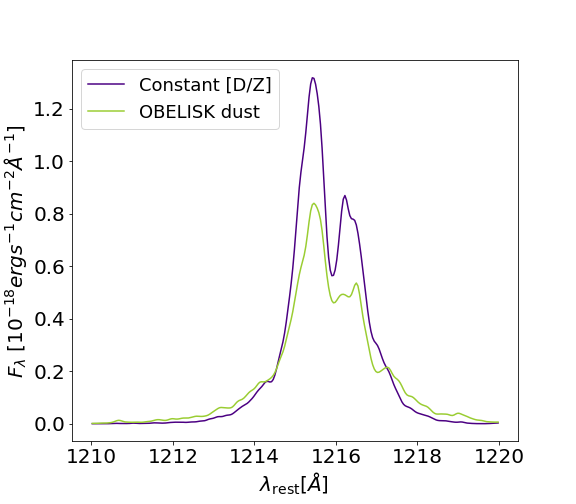}
    \caption{Comparison of the \lya{} profile of the same line of sight of galaxy ID 2899 for the two different dust models. The purple line is the observed spectrum for the constant $\mathcal{D}/Z$ model and the green line is the observed spectrum for the dust in \Obelisk.}
    \label{fig:dust_model_spectrum_comp}
\end{figure}

In Figure \ref{fig:dust_model_spectrum_comp}, we present a sample spectrum of a single line of sight of galaxy ID 2899 for both models. The green line shows the fiducial model from \Obelisk and the purple line shows the \citet{Laursen_2009} model in which the dust density is inferred from the local metallicity, based on LMC values, denoted as the constant $\mathcal{D}/Z$.
We see that the \lya{} flux observed from the run with the \Obelisk dust distribution is lower than in the case of the other dust model. This might be due to a difference in the spatial distribution of dust around the galaxy, which in the case of our fiducial dust model might be more concentrated around \lya{} emitting regions. However, the main features of the line are mostly unaffected by the choice of dust modeling. In particular, the position of each peak and the relative prominence of the blue peak are consistent between the two runs.
\begin{figure}
    \centering
    \includegraphics[width = 0.85 \linewidth]{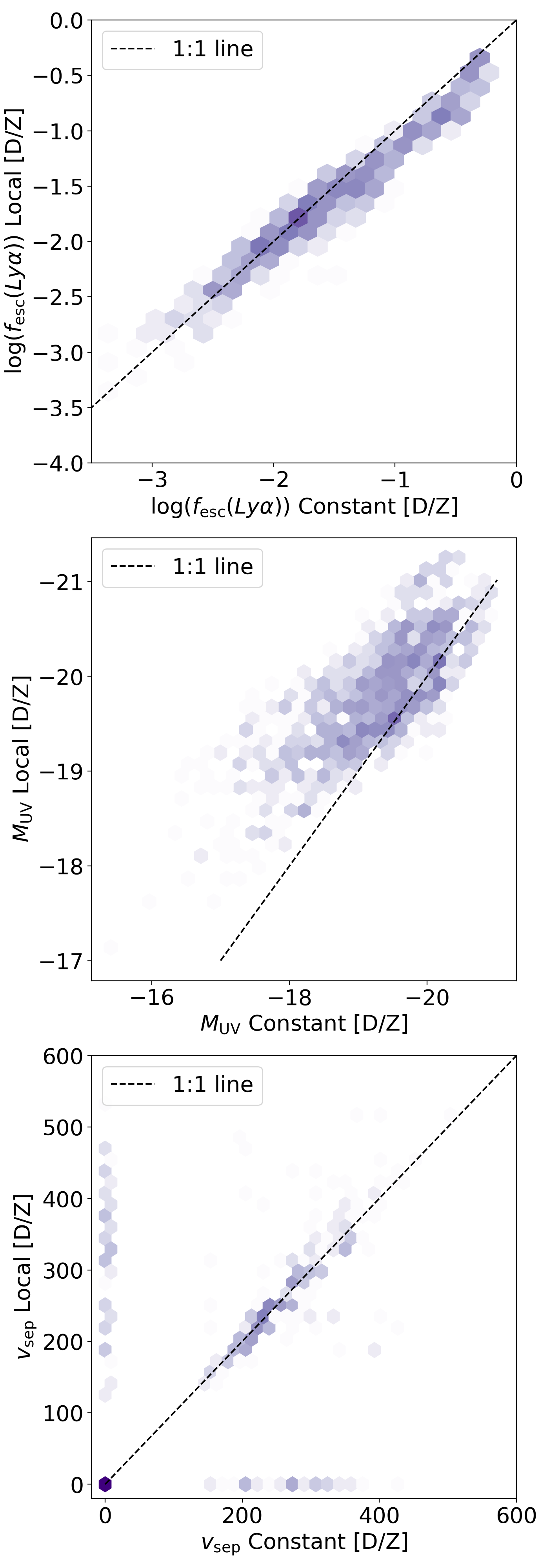}
    \caption{Comparison of log(\fescLya), UV magnitude, and Lyman-$\alpha$ peak separation (top, middle, and bottom panels, respectively) inferred from the synthetic spectra run with the two different dust models.
    In general, we see that the values are consistent with each other, as they do not substantially deviate from the one-to-one line, which is shown with the dashed black line in each panel.}
    \label{fig:dust_comparison}
\end{figure}

We now take a more quantitative view of the difference between the two sets of runs in Figure~\ref{fig:dust_comparison}. We compare the distribution of the \lya{} escape fraction $\fescLya$ (top), which, in simulations, depends on the absorption of photons by dust in the ISM, the absolute observed UV magnitude $\Muv$ (middle), and the \lya{} peak separation, $v_{\rm sep}$ (bottom). The values measured for our fiducial model, the dust measured from \Obelisk, are presented on the $y$ axis denoted by a local $\mathcal{D}/Z$, while the other model is denoted by a constant $\mathcal{D}/Z$. In each panel, the color indicates the density of points (i.e., the number of sight lines, with darker colors meaning more points), and the dashed black line shows the 1:1 line.

The top panel shows that the \lya escape fraction, \fescLya, is consistent between the two dust models, although there is some small deviation from the 1:1 line at $\fescLya \gtrsim 10\%$, where the constant $\mathcal{D}/Z$ dust model is higher. This is qualitatively consistent with the results of Figure~\ref{fig:dust_model_spectrum_comp}. 

In the second panel of Figure \ref{fig:dust_comparison}, we see that the observed UV magnitudes are generally brighter in the run with the \Obelisk dust distribution. Once again we can explain this by studying the difference in dust distribution between the two models. Indeed, assuming a constant $\mathcal{D}/Z$ over the whole ISM will lead to overestimating it in lower-density regions, where the $\mathcal{D}/Z$ in the simulation tends to be lower. Since UV photons escape more easily from these low-density regions, this will lead to higher attenuation, and so the model with constant $\mathcal{D}/Z$ will attenuate the UV more strongly, which explains the faint observed UV magnitudes. In Appendix \ref{ch:AppendixB} we show two maps of $\mathcal{D}/Z$ for our fiducial dust model for two of our galaxies to show that taking a constant $\mathcal{D}/Z$ would overestimate the dust content, especially in the low-density region.


Finally, in the last panel, we find that the measurement of the peak separation is mostly unaffected by the dust model used. This is consistent with the expectations from simpler expanding shell models: for example, \citet{Verhamme_2015} found that the position of the (red) peak is mostly independent of the dust optical depth of the shell.
However, for many lines of sight, we see that spectra presenting a double peak for a dust model present a single peak for the other. In most cases, this is due to one of the peaks becoming weaker, and therefore not significant enough to be identified by our peak finding routine, as is seen for example in Figure \ref{fig:dust_model_spectrum_comp}, where the red peak is much less marked when using the \Obelisk dust model.

Overall, the two dust models seem fairly consistent with each other. The \fescLya is consistent, with small differences at high escape fractions, and the UV magnitude is also fairly consistent, especially for brighter galaxies, while faint galaxies are about 1.5 dex from the 1:1 line. Lastly, the peak separation is consistent, when both peaks are detected in both profiles. This shows that although the choice of dust model can change the luminosity (in \lya or in UV) of the galaxy, the \lya spectral shape is mostly unaffected.
In the next section, we explore the usability of the shape of the \lya line as a diagnostic tool to understand the escape of ionizing radiation in our simulated LAEs. Motivated by the results of Figure~\ref{fig:dust_comparison}, we choose to focus on a single dust model, namely the one from \Obelisk, corresponding to a local $\mathcal{D}/Z$.

\section{\lya{} emission from simulated galaxies}\label{ch:Lya_properties}

In this section, we discuss the integrated \lya properties of our simulated sample and consider the diversity of \lya profiles resulting from the synthetic observations of our sample for the \Obelisk dust model, our fiducial model.

\begin{figure}
    \centering
    \includegraphics[width = 0.95 \linewidth]{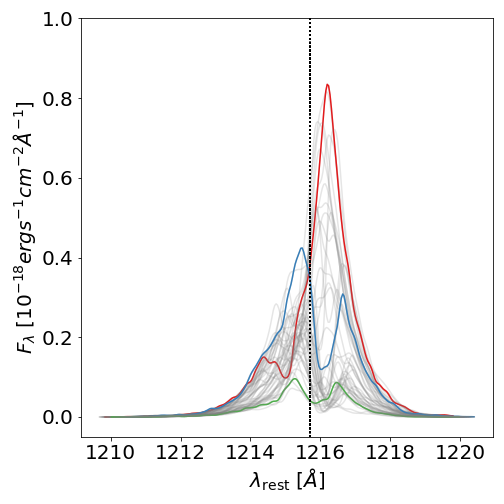}
    \caption{All 48 spectra of galaxy ID 8776, where the three colored spectra are meant to highlight the diversity of spectral shapes that are possible from the same galaxy.
    The dotted vertical black line is the Lyman-$\alpha$ wavelength at $z=6$.}
    \label{fig:all_spectra}
\end{figure}

We first present in Figure~\ref{fig:all_spectra} the emerging \lya profile of galaxy ID 8776 along 48 different lines of sight, each shown as a thin gray line. We highlight three of the lines of sight displaying representative but extremely different line shapes: in red a prototypical red peak, in green a (fairly attenuated) double peak, and in blue a line dominated by its blue peak. In general, we expect red-dominated profiles to trace outflows, and blue-dominated profiles to trace inflows \citep{Verhamme_2006}. This diversity is reminiscent of the results of, for example, \citet{blaizot2023simulating}, and corresponds to different orientations of both the galaxy and gas flows around it.  In Appendix \ref{ch:Appendix} we show the same figure for the other 12 galaxies. In addition to the different shapes of the line, the total \lya flux also varies with direction: the double peak highlighted in green in Figure \ref{fig:all_spectra}  appears to be substantially fainter than the spectrum with a red peak highlighted in red. This implies that the flux-selected sample of LAEs could miss some of the galaxies that happen to be observed from a direction where the flux is low, even if the galaxy emits a strong \lya line in other directions.
If we define LAEs as objects with a rest frame equivalent width of $EW_0 \gtrsim 20$\AA{} \citep{Ouchi_2020}, the galaxy in Figure \ref{fig:all_spectra} would result in an LAE in only 22 out of 48 lines of sight, so 46\% of the time. In the whole sample, this fraction varies from 31\% to 98\%, with a mean of 71\%.

We note that Figure \ref{fig:all_spectra} also shows that in many directions the spectra present a somewhat unexpectedly strong blue peak, which is the case for all of the galaxies in our sample.
There are several explanations for this: first of all, we do not model the attenuation from the IGM in this work, which would suppress our strong blue peaks. However, even at lower redshift, where IGM attenuation is less strong, observations tend to show a stronger red peak (e.g., \citealt{Verhamme_2016, Leclercq_2017, Izotov_2018, Orlitova_2018}).
A second possibility is that our strong blue peaks come from our choice of limiting the \lya transfer only up to the virial radius of each galaxy: \citet{blaizot2023simulating} have shown that transfer further than the virial radius tends to reduce the strength of the blue peaks. By excluding the gas between 1 and 3 $R_{\rm vir}$, we are (artificially) making this effect weaker. Indeed, in relatively low-mass systems like the LAEs we are modeling here, galactic outflows typically extend beyond the virial radius \citep{Mitchell_2017, Mitchell_2020}, and so by restricting the \lya transfer to the halo we might miss part of the outflow.

From each synthetic observation, we measure the peak separation, as detailed in Section~\ref{ch:mock_obs}, and using the intrinsic \lya luminosity of each galaxy (derived directly from the simulation), we measure the \lya escape fraction for each galaxy along each line of sight.
\begin{figure}
    \centering
    \includegraphics[width =0.95 \linewidth]{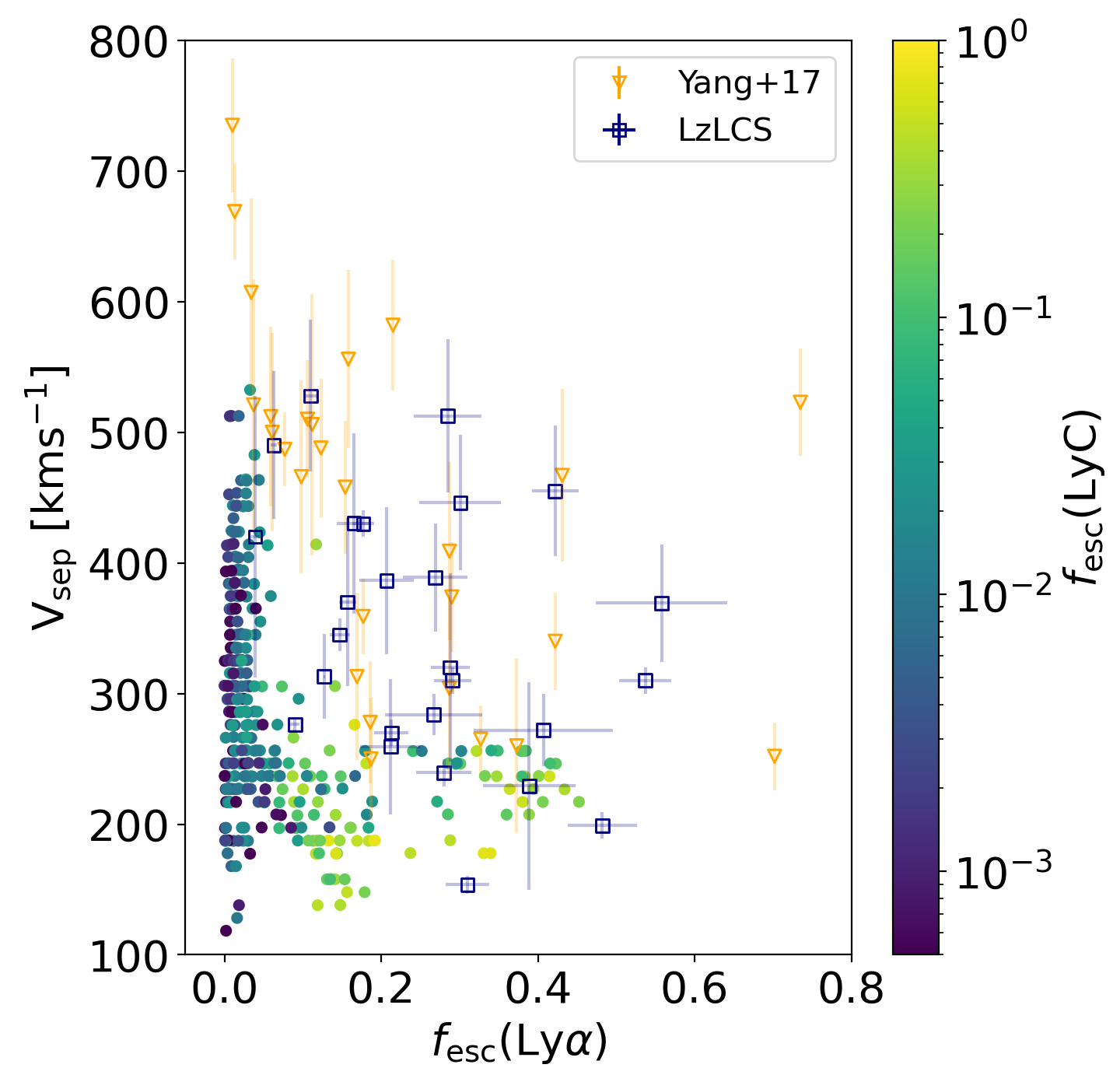}
    \caption{\lya{} peak separation against \lya{} escape fraction, color-coded by the LyC escape fraction. We also show observational data from \cite{Yang_2017} in the orange triangles and \cite{Flury_2022b} in the blue squares. Here, we are showing that although spectra with large peak separations (above 300 km/s) always have low LyC and \lya{} escape fractions, the contrary is not true, as we have many lines of sight with both low peak separation and escape fractions.}
    \label{fig:peak_sep_fesc}
\end{figure}
In Figure~\ref{fig:peak_sep_fesc}, we show the separation between the two peaks in the \lya{} profiles against the \lya{} escape fraction and we compare it to observational data from \cite{Yang_2017} and \cite{Flury_2022b}. We choose here to color-code our points by the LyC escape fraction along the same line of sight, \fescLyC, where lighter points correspond to a higher \fescLyC.
Here, we note that the (apparent) discreteness of the \lya peak separation values is completely artificial and comes from our choice of spectral resolution for the synthetic observations. All of the spectra were observed with the same resolution, in the same wavelength range, and were also smoothed in the same way, which leads to the discreteness in the values of the peak separation.

Overall, the figure displays a trend of peak separation decreasing with \fescLya, in good agreement with expectations from both earlier models (e.g., \citealt{Verhamme_2015}) and observations of low-z LAEs \citep{Yang_2017, Flury_2022b}: the sight lines with very high \fescLya tend to have low peak separations, and sight lines with peak separations above 300 km/s tend to have low \fescLya. However, we find that the low separation, low \fescLya corner of the figure is also populated: galaxies with $\fescLya \lesssim 10\%$ have a wide range of possible peak separations, in contrast both with observations and with shell models such as those of \citet{Verhamme_2015}, which predict a low \hi column density in lines of sight with low peak separation, and in contrast with observational data. We believe that this might be due to the clumpy nature of the galaxies modeled in this work. The \lya{} radiation can originate from different clumps in the galaxy, which might have different properties and different intrinsic velocities, which could contribute to the creation of multiple close peaks. For a detailed discussion on the nature of the difference between idealized models such as the picket fence model and results from radiative transfer in hydrodynamical simulations, we refer the reader to \cite{Mauerhofer_2021}.

\begin{figure}
    \centering
    \includegraphics[width = 0.95 \linewidth]{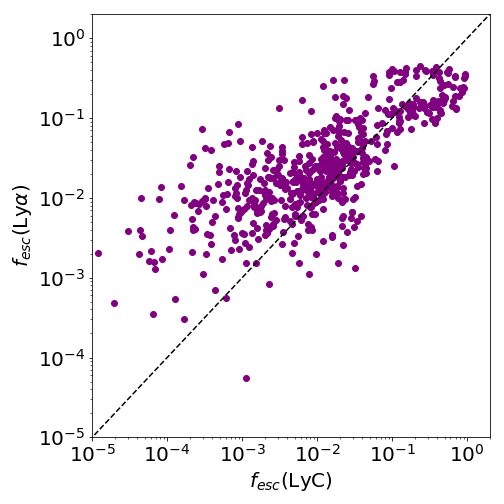}
    \caption{\fescLyC versus \fescLya. The black dashed line corresponds to the 1:1 relation.}
    \label{fig:lyc_vs_lya}
\end{figure}

A similar behavior is also seen for \fescLyC. High \fescLyC sight lines have low peak separation but low \fescLyC sight lines correspond to a large range of peak separations. This implies a relation between \fescLyC and \fescLya, which is confirmed by Figure \ref{fig:lyc_vs_lya}, which shows that an increase in \fescLya corresponds to an increase in \fescLyC, although with some scatter, especially at low \fescLyC. This relation is also expected from theory \citep[e.g.,][]{Dijkstra_2016, Kimm_2022, Maji_2022} and observations \citep[e.g.,][]{Flury_2022b, begley2023connecting}. A possible explanation for this scatter between \fescLyC and the \fescLya, and therefore also the \lya{} peak separation, could be the much more localized emission of the LyC radiation with respect to the \lya{} radiation, such as that observed in the Sunburst Arc by \citet{Kim_2023}. Indeed, they show that non-LyC-leaking regions of the Sunburst Arc still have substantial \fescLya of about 13\%, and it is therefore not unexpected to have very low \fescLyC while still being able to observe \lya{}.

\section{Estimate of \fescLyC}\label{ch:Estimate_lyc}

Several quantitative measurements based on the \lya line shape have been proposed in the literature to infer the escape of LyC radiation.
In this section, we will describe and analyze some of these methods: the \lya{} peak separation \citep[e.g.,][]{Izotov_2018}, the \lya{} central flux fraction proposed by \citet{Naidu_2022}, and the fraction of flux in the \lya trough compared to the continuum flux proposed by \citet{gazagnes_2020}.
Having explored the \lya line shape in the previous section, we will now assess how these proposed diagnostics perform for our simulated galaxy sample, and compare that to the ``true'' \fescLyC measured directly in the simulation.

\subsection{Peak separation, $v_{\rm sep}$}

From theoretical studies of spherical shell models, we expect a relation between the peak separation and the column density neutral gas in the line of sight \citep[e.g.,][]{Verhamme_2006, Orsi_2012, Verhamme_2015}, which is itself related to \fescLyC, where a higher column density corresponds to a lower escape fraction \citep{Verhamme_2016}. We therefore expect a relation between the \lya{} peak separation and \fescLyC. This relation was first shown in simulations by \citet{Verhamme_2016} and then calibrated by \citet{Izotov_2018}, using a set of low-redshift LyC-leaking galaxies known as “Green Peas” \citep{Cardamone_2009}. 
This particular set of galaxies was used as they are confirmed LyC leakers, at a redshift high enough that the Milky Way's neutral gas would not absorb the LyC radiation, but low enough to avoid LyC absorption by the neutral IGM and to be detected by the Cosmic Origin Spectrograph on the \textit{Hubble Space Telescope (HST)}. The Green Pea galaxies were selected for their high $O_{32}$. Their SFRs are comparable to the most massive galaxies in our sample but are brighter than our sample, with UV magnitudes around -19.5 to -20.5. The stellar masses of the Green Peas are also comparable with those of the galaxies in our sample.
The relation found by \citet{Izotov_2018} is
\begin{equation}\label{eq:peak_sep}
    f_{esc}(\textrm{LyC}) = \frac{3.23 \times 10^4}{V_{sep}^2} - \frac{1.05 \times 10^2}{V_{sep}} + 0.095
.\end{equation}
This is empirically derived and breaks down at $v_{\rm sep} < 140 \textrm{ km/s}$, where it gives unphysical values of \fescLyC$>1$. It can nonetheless be used to infer the escape fraction of ionizing radiation, or a lower limit in the case of low peak separations, and has already been applied at high redshift \citep[e.g.,][]{Meyer_2020}. 
The anticorrelation between peak separation and the LyC escape fraction found by \citet{Izotov_2018} was also found in other observations of low-redshift LyC leakers \citep{Verhamme_2016, Flury_2022b, Izotov_2022} and several radiation hydrodynamic simulations of cloud-scale LyC escape \citep{Kimm_2019, Kakiichi_2021}.

\begin{figure}
    \centering
    \includegraphics[width =  \linewidth]{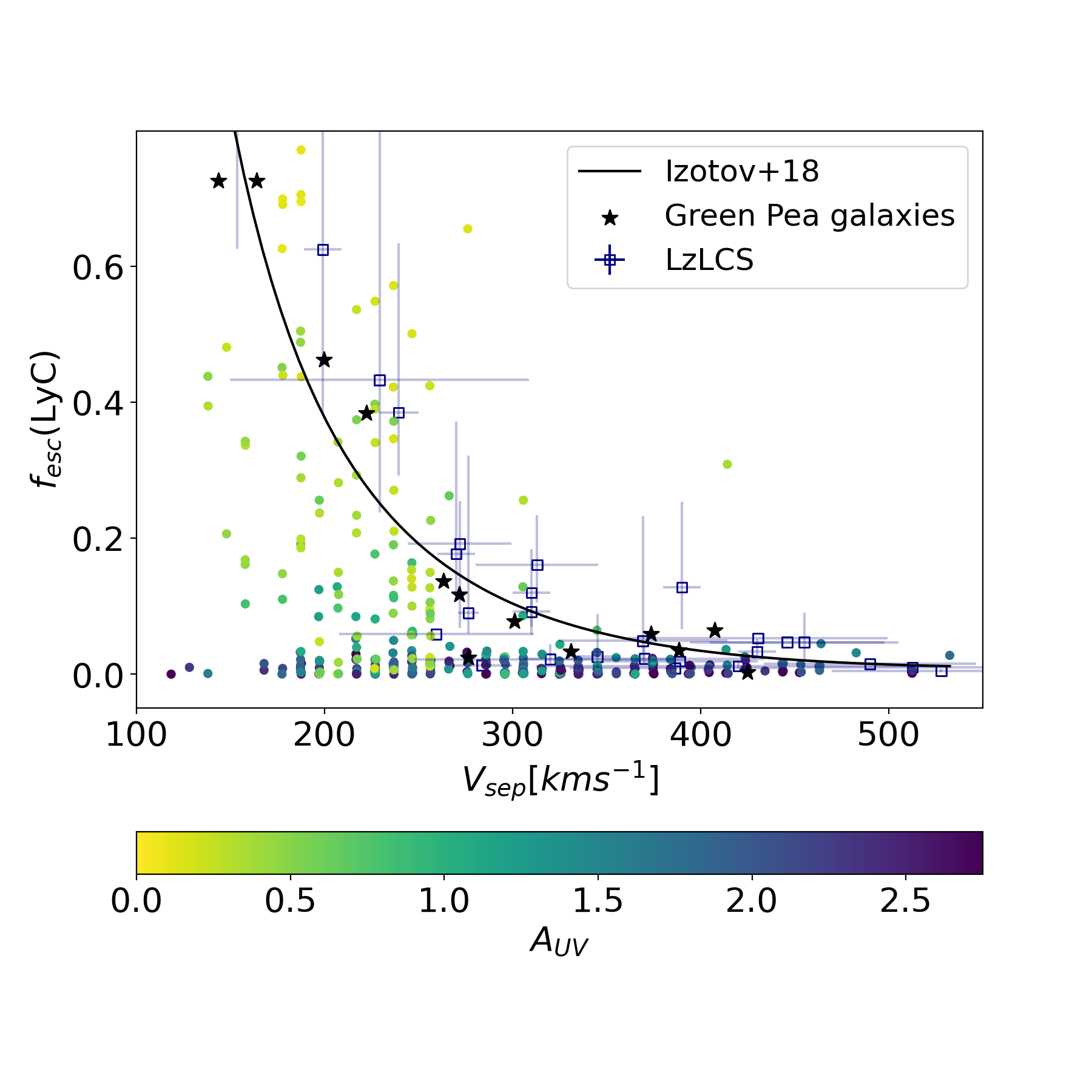}
    \caption{Comparison of our data, color-coded by the dust attenuation in the UV of each line of sight, with the relation found by \citet{Izotov_2018}, shown with the black line. The Green Pea galaxies that the relation was calibrated on are shown with black stars. The more recent measurements of peak separations of low-redshift galaxies by \cite{Flury_2022b} are shown with empty blue squares.}
    \label{fig:peak_sep}
\end{figure}
In Figure \ref{fig:peak_sep}, we show where our simulated data lie in the \fescLyC to $v_{\rm sep}$ plane, color-coded by the dust attenuation in the UV for each line of sight, and compare it with the best-fit relation from \citet{Izotov_2018}. We also show the Green Pea galaxies this relation was calibrated on with the black stars and the LzLCS results in the empty blue squares.

Overall, there is a reasonable agreement between the simulation and the observations, although with a large scatter: \Obelisk reproduces the trend that a larger peak separation indicates a lower \fescLyC. Nevertheless, we also note a large number of sight lines with low peak separation and low \fescLyC, suggesting that the observed relation is not fully representative of our simulated sample. This is in agreement with the results from the SPHINX$^{20}$ simulation \citep{choustikov2024great}. We find 97 sight lines with $0<v_{\rm sep}<250$ km/s and \fescLyC$<0.1$, or 15\% of the total sight lines, which is a range not yet probed by observations, although new \textit{HST} observations were recently designed to probe this parameter space (\citealt{Leclercq_2022}, GO-17153, PI Leclercq). We find that these results are consistent with the work of \citet{Kimm_2019}, who also find a population of sight lines around a simulated molecular cloud that have low peak separation but low \fescLyC.

Interestingly, points with low dust attenuation (in yellow) tend to align better with the \citet{Izotov_2018} relation, while sight lines with high dust attenuation tend to depart from it. Using data from LzLCS, \citet{Saldana_Lopez_2022} show that it is necessary to account for dust attenuation in the indirect diagnostics of LyC escape: our numerical results seem to confirm this.
This, however, does not mean that dust absorbs LyC radiation: indeed, in general we do not see more than a 1\% difference in the \fescLyC between runs with or without dust. Instead, our interpretation is that as the amount of dust in a given line of sight scales with the amount of hydrogen, sight lines with higher dust attenuation are also those with higher \hi column density. Therefore, lines of sight with higher dust attenuation should correspond to a lower LyC escape fraction. Here, we would like to note that we find lines of sight with high dust attenuation but low peak separation, which may seem unexpected, but these are the same lines of sight as that in the bottom left corner of Figure \ref{fig:peak_sep_fesc}, which has a low peak separation and low \fescLya. We therefore believe that the explanation that was given in Section \ref{ch:Lya_properties} about the clumpy nature of our galaxies applies here as well. 

Another possible explanation for this discrepancy between the simulated data and the Green Peas is that Green Pea galaxies might be intrinsically different from the galaxies in our sample. Green Pea galaxies are slightly brighter, with $-19.5 < M_{UV} < -20.5$, and were selected for their strong oxygen line ratio, $O_{32}$. Similarly, the LzLCS galaxies are mostly brighter than ours, although their $v_{\rm sep}$ sample has magnitudes ranging between $-18.5 < M_{UV} < -21$.
Both observational samples also comprise galaxies usually found in low-density environments at low redshift, while \Obelisk simulates a high-density environment at high redshift, so it could also be an environmental effect.

\subsection{Central flux fraction}

The amount of flux leftover at the line center is expected to be a good indicator of both the \lya{} and the LyC escape fraction, as it should correspond to \lya photons that are able to escape at the systemic redshift and have therefore been scattered only a few times. These photons would have escaped through clear sight lines in a partially transparent ISM, where we also expect high \fescLyC. If there are many clear sight lines, we can expect a third, central peak, making the central flux fraction method derived by \cite{Naidu_2022} particularly useful when dealing with \lya profiles that have more than two peaks.
\citet{Naidu_2022} define the central flux fraction, $f_{\rm cen}$, as the fraction of \lya{} flux within $\pm 100$ km s$^{-1}$ of the systemic velocity:
\begin{equation}
    \textrm{Central flux fraction }(f_{\rm cen}) = \frac{\textrm{Ly}\alpha \textrm{ flux at}\pm \textrm{100 km s}^{-1}}{\textrm{Ly}\alpha \textrm{ flux at}\pm \textrm{1000 km s}^{-1}}
,\end{equation}
and show that in their sample of 25 literature LyC leakers at low redshift ($z\approx0.3-4$) a central flux fraction above 10\% corresponds to a high \fescLyC ($f_{\mathrm{esc}}>20\%$), although some high \fescLyC galaxies (two out of nine) are missed with this method. The low \fescLyC galaxies (\fescLyC<5\%) are instead selected as having both $f_{\rm cen}<10\%$ and $v_{sep}>375$ km s$^{-1}$. It must also be noted that the choice of $\pm 100$ km s$^{-1}$ as the central flux is resolution-dependent and might therefore affect our estimates, as our resolution is higher.

\begin{figure}
    \centering
    \includegraphics[width =0.95 \linewidth]{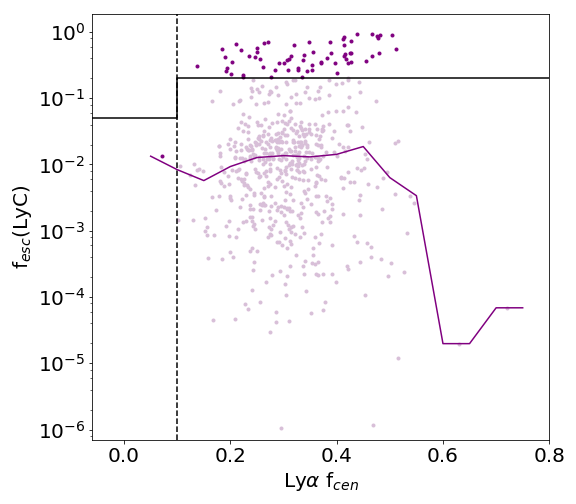}
    \caption{Comparison of our data with the central flux fraction method derived by \citet{Naidu_2022}. The dashed black line here is the divide between leakers and non-leakers. The points in purple are the points from our sample that fit these criteria and the points in pastel are those that do not. We also show the running median of our sample here with the solid purple line.}
    \label{fig:central_flux}
\end{figure}
In Figure \ref{fig:central_flux}, we show \fescLyC as a function of the central flux fraction for our sample, with the divide from \citet{Naidu_2022} as the dashed black line: the point on the left of the vertical dashed line would be expected to have $\fescLyC < 10\%$ solely based on its $f_{\mathrm{cen}}$, while points to the right of the line would have $\fescLyC > 20\%$. In the figure, the dark purple points are those in our sample that fit the criteria from \citet{Naidu_2022}, and the pastel points are those that do not. We also show the running median with the solid purple line, which clearly lies almost an order of magnitude below the criterion of \citet{Naidu_2022}, indicating that for any given $f_{\rm cen}$ most of the sight lines in our sample would be classified as non-leakers. This implies that for the typical sight line of our simulated LAEs, $f_{\rm cen}$ is not a good predictor of \fescLyC. Here, we would like to note that the upturn in \fescLyC at high \lya{} $f_{\rm cen}$ is purely due to low number statistics, as the median in the last two bins is only calculated on one data point.

\subsection{$F_{\rm trough}/F_{\rm cont}$}

As was already described in the previous section, we expect leftover \lya{} flux at the line center to be an indicator of \fescLyC. The amount of leftover flux can be estimated as the fraction of flux at the line center over the continuum flux, since the presence of flux at the trough of the \lya{} profile above the continuum should indicate the escape through transparent channels. This ratio could therefore be a reliable indicator of properties of the ISM that favor LyC leakage. \cite{gazagnes_2020} derived a predictor for \fescLyC based on the ratio between the flux at the trough of the \lya{} spectrum and the continuum flux, calibrated on a sample of 22 star-forming galaxies with spectroscopy in the rest-frame UV. The relation is as follows:
\begin{equation}
    f_{\textrm{esc}}(\textrm{LyC}) = (0.032 \pm 0.006)\times \frac{\textrm{F}_{\textrm{trough}}}{\textrm{F}_{\textrm{cont}}} - 0.032 \pm 0.053
.\end{equation}

\begin{figure}
    \centering
    \includegraphics[width =0.95 \linewidth]{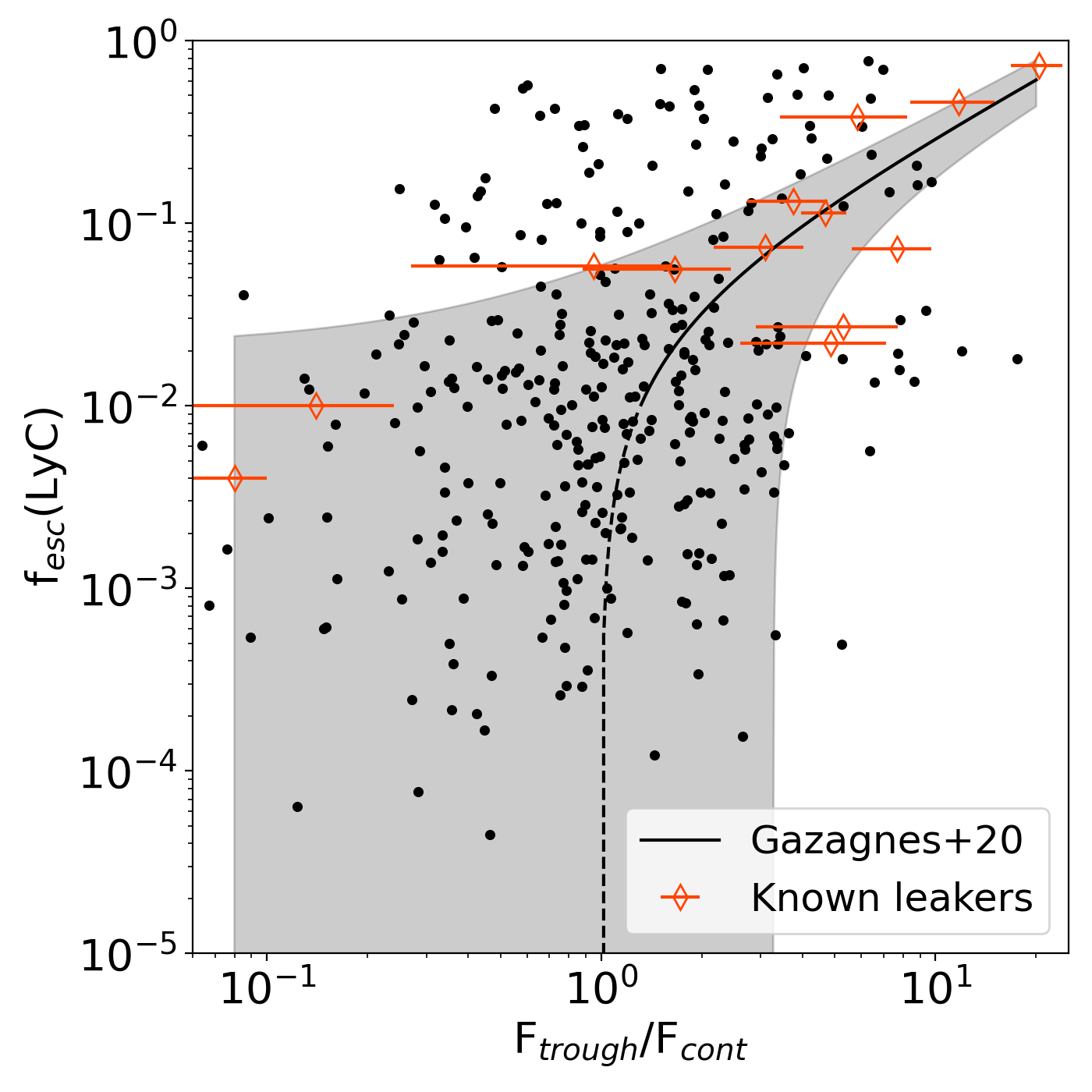}
    \caption{Comparison of our data with the relation found by \citet{gazagnes_2020} between \fescLyC and the fraction of the flux at the trough of \lya{} and the continuum flux. The shaded gray area represents the error on the relation. The empty red diamonds are the known leakers used by \citet{gazagnes_2020} to derive this relation. Many of our points fall on their relation but mostly for low escape fractions, whereas high escape fractions are often underestimated.}
    \label{fig:Ftrough_fcont}
\end{figure}
In figure \ref{fig:Ftrough_fcont}, we show our \fescLyC as a function of the $\frac{F_{\mathrm{trough}}}{F_{\mathrm{cont}}}$ derived from the synthetic spectra, with the relation found by \citet{gazagnes_2020} shown with the black line, their uncertainty in gray, and the leakers used to derive this relation as empty red diamonds. At \fescLyC<1\% we show the relation in a dashed line to indicate that the relation was extrapolated. Although many of our lines of sight fall within the uncertainty of the relation, this is mostly the case for low LyC escape fractions.
From Figure \ref{fig:Ftrough_fcont} we can see that these high escape fraction lines of sight are actually often underestimated by this relation: we find sight lines with $\fescLyC \gtrsim 10\%$ even at $F_{\rm trough}/F_{\rm cont} < 1$ (58 out of a total 452 sight lines have $F_{\rm trough}/F_{\rm cont} < 1$, so 13\%), and 45\% of points with high (>4) $F_{\rm trough}/F_{\rm cont}$ have low \fescLyC. This large scatter between the \fescLyC and $F_{\rm trough}/F_{\rm cont}$ could be due to the presence of dust in the line of sight, which affects \lya{} and LyC emission differently, and by the fact that \lya{} could scatter back into the line center, which could add flux at $F_{\rm trough}$ in low $f_{\rm esc}$ sight lines.

\subsection{Adding dust attenuation}

\begin{figure}
    \centering
    \includegraphics[width = 0.95 \linewidth]{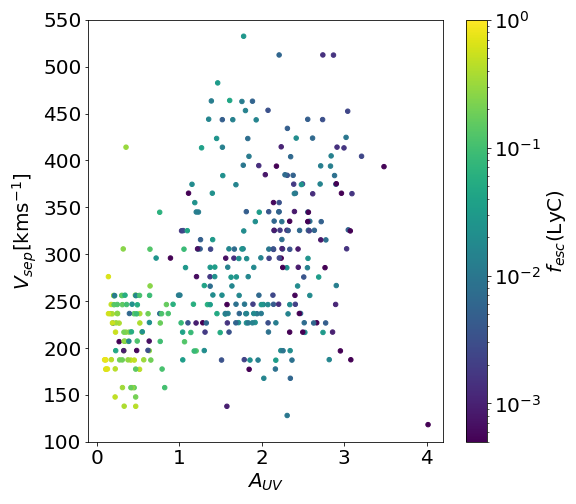}
    \caption{Peak separation against UV dust attenuation, color-coded by \fescLyC. We find high escape fractions for low peak separations, as was predicted by \citet{Izotov_2018}, but only in addition to low UV attenuation.}
    \label{fig:A_UV}
\end{figure}

None of the estimators presented in this section are sufficient to predict the range of escape fractions we see in our data. However, in Figure \ref{fig:peak_sep} we see a clear trend where lines of sight with higher $A_{UV}$ tend to have lower \fescLyC than lines of sight with lower $A_{UV}$. A similar trend has already been shown both in simulations \citep[e.g.,][]{Ma_2020, Yeh_2023} and in observations of low-redshift LyC leakers \citep{Chisholm_2022, Saldana_Lopez_2022}.
We investigate this behavior further in our sample by looking at the peak separation as a function of the dust attenuation, color-coded by the LyC escape fraction, shown in figure \ref{fig:A_UV}.
Here, we can clearly see that the lines of sight with both low peak separation and low dust attenuation are those that have a high LyC escape fraction. In our sample, 68\% of sight lines with $v_{\rm sep}<300$ and $A_{UV}<1$ have \fescLyC$>$10\%, while outside of this selection only 1.6\% of sight lines have \fescLyC$>$10\%. Moreover, both quantities can be found with observations, meaning that it should be possible to get better estimates for LyC escape fractions from observations of \lya{} profiles if the dust attenuation is also taken into account. We want to stress again that our values for \fescLyC are not strongly dependent on the presence of dust in the run, and therefore the relation we see is due to the amount of \hi in a sight line, which is a quantity traced by dust attenuation.

\section{Conclusions}\label{ch:conclusions}

In this work, we have studied a sample of 13 galaxies from the \Obelisk simulation at redshift z=6 by performing synthetic observations of the \lya{} line. Our sample was chosen to represent a flux-selected
sample targeting faint Lyman-break galaxies corresponding to typical LAEs at the end of the EoR.

We have shown that a large variety of \lya{} profiles can emerge from a single gas distribution, highlighting the challenges that come with the study of this line when it is used to trace intrinsic galaxy properties. We have also shown that although from theory we expected a low \lya{} peak separation to correlate with a high LyC escape fraction \citep{Verhamme_2016, Izotov_2018}, this does not hold for our sample, where we see a significant number of lines of sight with a low peak separation but also a low \lya{} and LyC escape fraction. 

We have also determined the impact that the choice of dust model has on our results by comparing the same properties found with two different dust models. Although there is a deviation from the 1:1 line in the case of the \lya{} escape fraction and the UV magnitude, the properties seem robust during a change in the dust model. We have investigated the reason for this deviation and we hypothesize that it could be due to the \Obelisk dust being more concentrated around star-forming regions.

Lastly, we compared methods of determining the LyC escape fraction calibrated with observations with the LyC escape fraction that we find from the simulation. 
We first considered the peak separation method, derived by \citet{Izotov_2018} based on Green Pea galaxies. This method relates the peak separation of the \lya{} line to the LyC escape fraction. Many of our points do not follow the relation, although here we are able to see a trend between the points that follow the relation and those that do not, which is the amount of dust attenuation in the UV in a given line of sight.
We have also compared our values with the central flux fraction method derived by \citet{Naidu_2022} and shown that many of our lines of sight have high $f_{\rm cen}$ but low \fescLyC, unlike their prediction. We show that the running median of our \fescLyC sample is about an order of magnitude lower than the 20\% value quoted by \citet{Naidu_2022} as the escape fraction expected for galaxies with  $f_{\rm cen}>10\%$.
We lastly compared our values with the $\frac{F_{\rm trough}}{F_{\rm cont}}$ method, designed by \citet{gazagnes_2020}. In this case, most of our lines of sight fall within the uncertainty but mostly at a low escape fraction. Most high-escape-fraction lines of sight, which are the most interesting in a study of the EoR, seem to be underestimated by this method. 
Following the trend we saw with the dust attenuation, we have looked at the relation between UV dust attenuation, peak separation, and LyC escape fraction. We have seen that lines of sight with low peak separation and UV attenuation are those with the highest LyC escape fractions. Overall, we see that adding a dust attenuation parameter seems to better estimate the LyC escape fractions, and in general we conclude that using multiple parameters is necessary to properly estimate \fescLyC.

\begin{acknowledgements}
We thank the referee for the useful comments on an earlier version of the manuscript. We thank Sophia Flury for providing the LzLCS data. MT, VM and PD acknowledge support from the NWO grant 016.VIDI.189.162 (``ODIN"). PD warmly thanks the European Commission's and University of Groningen's CO-FUND Rosalind Franklin program for support.
This work has received funding from the Swiss State Secretariat for Education, Research and Innovation (SERI) under contract number MB22.00072, as well as from the Swiss National Science Foundation (SNSF) through project grant 200020\_207349.
The Cosmic Dawn Center (DAWN) is funded by the Danish National Research Foundation under grant DNRF140.
\end{acknowledgements}

%
%
\bibliographystyle{aa}
\bibliography{aanda}
\onecolumn

\begin{appendix}
\section{\lya{} Profiles}\label{ch:Appendix}
\begin{figure}[h!]
    \centering
    \captionsetup{justification=centering}
    \includegraphics[width=\textwidth]{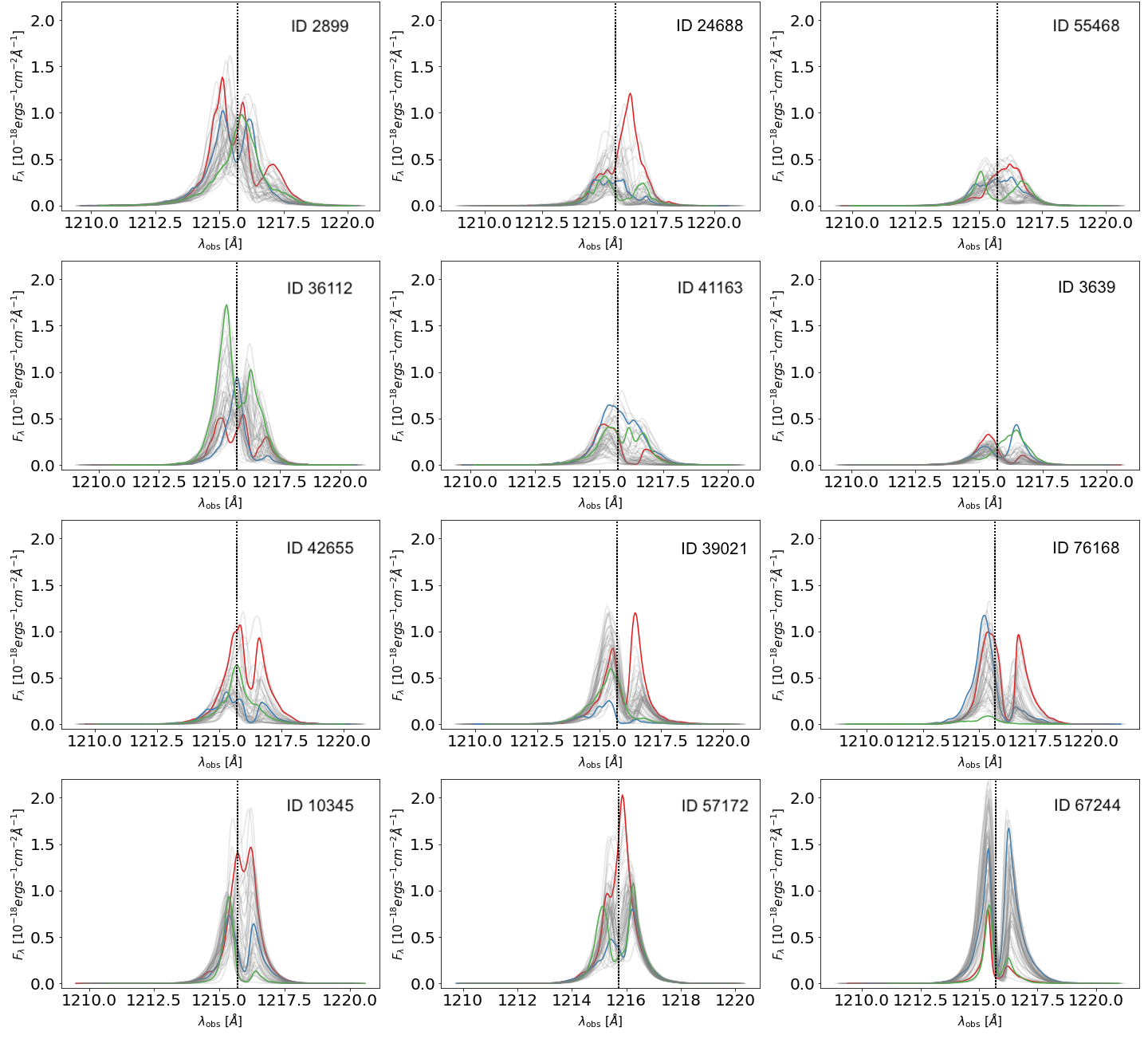}
    \caption{Similar to Figure \ref{fig:all_spectra} but for the rest of the galaxy sample}
    \label{fig:enter-label}
\end{figure}
\newpage
\section{Dust-to-metal ratio in \Obelisk}\label{ch:AppendixB}
\begin{figure*}[h!]
    \centering
    \captionsetup{justification=centering}
    \includegraphics[width=0.48\textwidth]{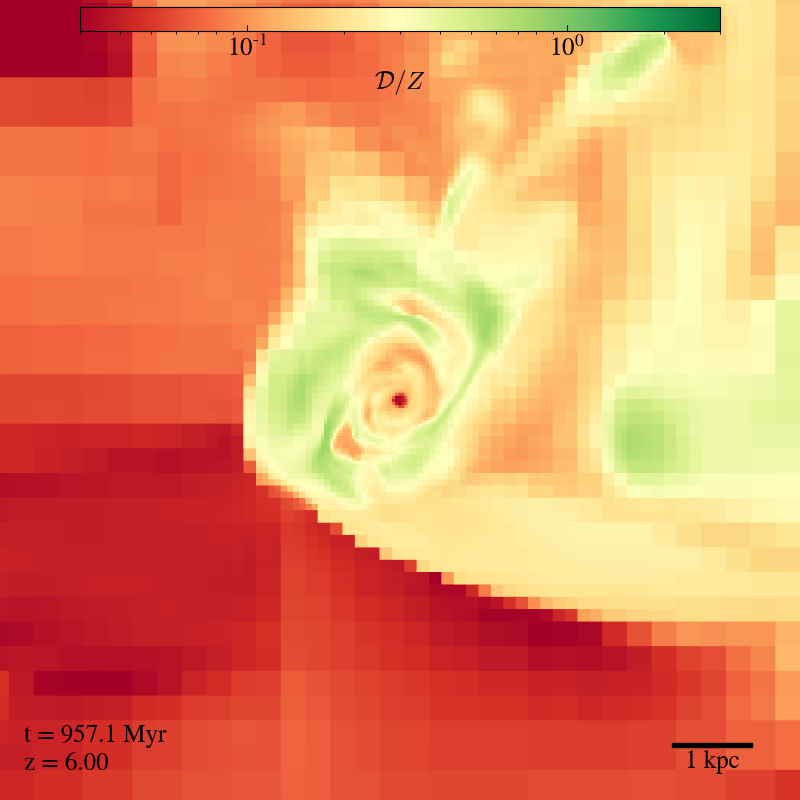}
    \includegraphics[width=0.48\textwidth]{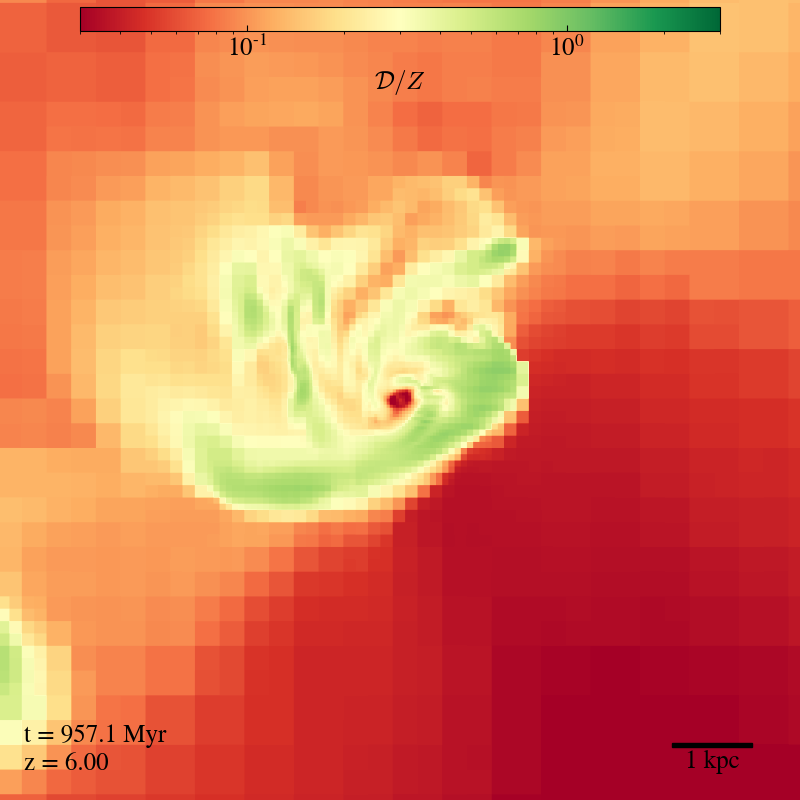}
    \caption{Dust-to-metal ratio distribution in \Obelisk for galaxies ID 2899 (left) and ID 55468 (right). Our constant $\mathcal{D}/Z$ model would overestimate the amount of dust in all the red and orange pixels in the figure, as the color bar is centered with yellow as the $\mathcal{D}/Z$ for the LMC}
    \label{fig:DtoZ}

\end{figure*}

\end{appendix}

\end{document}